\documentclass[11pt]{article}
\pdfoutput=1
\usepackage{jheppub}
\usepackage[OT1]{fontenc}
\usepackage{slashed}

\def\beq{\begin{equation}}
\def\eeq{\end{equation}}
\def\bec{\begin{center}}
\def\eec{\end{center}}

\DeclareMathOperator{\re}{Re}

\title{\bf The triple Higgs coupling: A new probe of low-scale seesaw
  models}

\author{Julien Baglio$^*_{}$,}
\affiliation{$^*_{}$ Institute for Theoretical Physics,  University of
  T\"ubingen, Auf der Morgenstelle 14, 72076~T\"ubingen, Germany}
\author{C\'edric Weiland$^\dagger_{}$}
\affiliation{$^\dagger_{}$ Institute for Particle Physics Phenomenology, Department
  of Physics, Durham University,  South Road, Durham DH1 3LE, United
  Kingdom}
\emailAdd{julien.baglio@uni-tuebingen.de}
\emailAdd{cedric.weiland@durham.ac.uk}

\abstract{The measure of the triple Higgs coupling is one of the major
  goals of the high-luminosity run of the CERN Large Hadron Collider
  (HL-LHC) as well as the future colliders, either leptonic such as
  the International Linear Collider (ILC) or hadronic such as the 100
  TeV Future Circular Collider in hadron-hadron mode (FCC-hh). We have
  recently proposed this observable as a test of neutrino mass
  generating mechanisms in a regime where heavy sterile neutrino
  masses are hard to be probed otherwise. We present in this article a
  study of the one-loop corrected triple Higgs coupling in the inverse
  seesaw model, taking into account all relevant constraints on the
  model. This is the first study of the impact on the triple Higgs
  coupling of heavy neutrinos in a realistic, renormalizable neutrino
  mass model. We obtain deviations from the Standard Model as large as
  to $\sim +30\%$ that are at the current limit of the HL-LHC
  sensitivity, but would be clearly visible at the ILC or at the
  FCC-hh.}

\arxivnumber{1612.06403}
\preprint{IPPP/16/116}

\allowdisplaybreaks[2]

\begin{document}

\thispagestyle{empty}
\def\thefootnote{\fnsymbol{footnote}}
\setcounter{footnote}{1}

\setcounter{page}{0}
\maketitle
\flushbottom

\def\thefootnote{\arabic{footnote}}
\setcounter{footnote}{0}
%
%
\section{Introduction}
The CERN Large Hadron Collider (LHC) was home to one of the biggest
discoveries in particle physics with the observation of a Higgs boson
with a mass of around 125~GeV in
2012~\cite{Aad:2012tfa,Chatrchyan:2012ufa}, thanks to the data
collected in Run 1 at 7 and 8 TeV. The Higgs boson is the remnant of
the electroweak symmetry-breaking (EWSB)
mechanism~\cite{Higgs:1964ia,Englert:1964et,Higgs:1964pj,Guralnik:1964eu}
that generates the masses of the other fundamental particles and
unitarizes the scattering of weak
bosons~\cite{Cornwall:1973tb,LlewellynSmith:1973yud}. The Run 2 data
collected in 2015 and 2016 at a center-of-mass energy of 13 TeV still
displays a compatibility of this Higgs boson with the Standard Model
(SM) hypothesis; nevertheless we know that the SM cannot be the
ultimate theory. In particular the observation of neutrino
oscillations, confirmed in 1998 at
Super-Kamiokande~\cite{Fukuda:1998mi}, implies that neutrinos are
massive, which cannot be explained in the SM framework and thus calls
for an extension of the SM. One of the simplest possibilities to
explain the non-zero neutrino masses and mixing is to add fermionic
gauge singlets that will play the role of right-handed neutrinos. The
addition of these heavy sterile neutrinos leads to the type I seesaw
model and its various
extensions~\cite{Minkowski:1977sc,Ramond:1979py,GellMann:1980vs,Yanagida:1979as,Mohapatra:1979ia,Schechter:1980gr,Schechter:1981cv,Mohapatra:1986aw,Mohapatra:1986bd,Bernabeu:1987gr,Pilaftsis:1991ug,Ilakovac:1994kj,Akhmedov:1995ip,Akhmedov:1995vm,Barr:2003nn,Malinsky:2005bi}.
A very recent study summarizes the possible direct detection
possibilities and indirect tests for heavy sterile neutrinos at lepton-lepton,
proton-proton and lepton-proton colliders~\cite{Antusch:2016ejd}, see
also references therein.

In a recent article~\cite{Baglio:2016ijw} we have presented the triple
Higgs coupling $\lambda_{HHH}^{}$ as a new observable to test neutrino
mass generating mechanisms in a regime of mass difficult to probe otherwise. The
measure of $\lambda_{HHH}^{}$ is one of the main goals of the
high-luminosity run of the LHC (HL-LHC) as well as of the future
colliders, such as the electron-positron International Linear Collider
(ILC)~\cite{Baer:2013cma} or the Future Circular Collider in
hadron-hadron mode (FCC-hh), a potential 100 TeV $pp$ collider (for
the Higgs studies see reviews in
refs.~\cite{Arkani-Hamed:2015vfh,Baglio:2015wcg,Contino:2016spe}). It
would be a direct probe of the shape of the scalar potential that
triggers EWSB. Any deviation of this coupling from the SM prediction
is then welcomed to unravel new physics. In ref.~\cite{Baglio:2016ijw}
the study of neutrino effects on $\lambda_{HHH}^{}$ was done in the
context of a simplified model with the SM plus one heavy Dirac
neutrino. It was found that effects as large as $+30\%$ at one-loop
could be obtained, at the limit of the currently foreseen $\sim
35\,\%$ sensitivity that the HL-LHC will have to the SM triple Higgs
coupling, when combining ATLAS and CMS data~\cite{CMS-PAS-FTR-15-002},
but clearly measurable at the ILC~\cite{Fujii:2015jha} or the
FCC~\cite{He:2015spf}. A comprehensive study in a realistic and
renormalizable model of neutrino masses was still left to be done.

In this article, we fill the gap and present the first analysis of
Majorana neutrino effects on $\lambda_{HHH}^{}$. We work within the
inverse seesaw (ISS)
model~\cite{Mohapatra:1986aw,Mohapatra:1986bd,Bernabeu:1987gr}, a
renormalizable low-scale seesaw model generating neutrino masses. After
taking into account all relevant constraints, we obtain effects that
can be as large as a $\sim +30\%$ increase of $\lambda_{HHH}^{}$,
similar to the effects that we found in our previous
article~\cite{Baglio:2016ijw} using a simplified model. In the case of
the ISS model, more heavy neutrinos are present, enhancing the effects
as we expected, but the constraints on the model are stronger,
reducing the end-effect back to the simplified model
expectations. This can be clearly measurable at the ILC and at the
FCC-hh and is at the limit of the currently foreseen sensitivity of
the HL-LHC.

This article is organized as follows. In Section~\ref{sec:model} we
present the ISS model as well as the theoretical and experimental
constraints that we consider. We give the technical details of our
calculation in Section~\ref{sec:calc} and present the numerical
analysis of the ISS one-loop corrections to $\lambda_{HHH}^{}$ in
Section~\ref{sec:pheno}. A short conclusion is given in
Section~\ref{sec:conc}. We present the details of the parameterization
adopted for the light neutrino mass matrix in Appendix~\ref{app:NLO}
and the analytical expressions of the one-loop corrections involving
the neutrinos are collected in Appendix~\ref{app:HHH}.

%
%
\section{Model and constraints}
\label{sec:model}

While our calculation and the analytical results presented in
Section~\ref{sec:calc} are applicable to all models with extra
fermionic gauge singlets and Majorana neutrinos like the type I
seesaw~\cite{Minkowski:1977sc,Ramond:1979py,GellMann:1980vs,Yanagida:1979as,Mohapatra:1979ia,Schechter:1980gr,Schechter:1981cv}
or the linear
seesaw~\cite{Akhmedov:1995ip,Akhmedov:1995vm,Barr:2003nn,Malinsky:2005bi}, 
we will focus in this work on the inverse seesaw (ISS) model for
illustrative purposes. After introducing the model and the different
parameterizations used to reproduce neutrino oscillations data, we will
present the theoretical and experimental constraints considered in our
study.

\subsection{The inverse seesaw model}
\label{sec:ISS}

One particular variant of the type I seesaw is the ISS
model~\cite{Mohapatra:1986aw,Mohapatra:1986bd,Bernabeu:1987gr} which 
has very interesting characteristics leading to a rich
phenomenology. In the ISS model the suppression mechanism that
guarantees the smallness of neutrino masses is the introduction of a
slight breaking of lepton number in the singlet sector (composed of
right-handed neutrinos $\nu_R^{}$ and new gauge singlets $X$ with
opposite lepton number), in the form of a small Majorana mass
$\mu_X^{}$ for the $X$ singlets, compared to the electroweak scale
$v\sim 246$~GeV. This allows for large Yukawa couplings compatible
with a low (TeV or even lower) mass for the seesaw mediators, contrary
to the seesaw model of type I for example, where the mediators have a
mass of the order of the GUT scale or the Yukawa couplings are very
small.

In the inverse seesaw, the additional terms to the SM Lagrangian are
\begin{equation}
  \label{LagrangianISS}
  \mathcal{L}_\mathrm{ISS} = - Y^{ij}_\nu \overline{L_{i}}
  \widetilde{\Phi} \nu_{Rj} - M_R^{ij} \overline{\nu_{Ri}^C} X_j -
  \frac{1}{2} \mu_{X}^{ij} \overline{X_{i}^C} X_{j} + h.c.\,,
\end{equation}
where $\Phi$ is the Higgs field and $\widetilde \Phi=\imath \sigma_2
\Phi^*$, $i,j=1\dots 3$, $Y_\nu$ and $M_R$ are complex matrices and
$\mu_{X}$ is a complex symmetric matrix whose norm is taken to be
small since lepton number is assumed to be nearly conserved. In this
work, we do not consider a possible Majorana mass term for the
right-handed neutrinos $\nu_R$ since this extra parameter is not
relevant to our study. It would only induce negligible corrections to
the heavy neutrino masses and the observable that we consider
conserves lepton number. Assuming 3 pairs of $\nu_R$ and $X$, the
$9\times 9$ neutrino mass matrix reads after electroweak symmetry
breaking in the basis $(\nu_L^C\,,\;\nu_R\,,\;X)$,
\begin{equation}
  \label{ISSmatrix}
  M_{\mathrm{ISS}} =
  \left(
    \begin{array}{c c c} 0 & m_D & 0 
      \\ m_D^T & 0 & M_R 
      \\ 0 & M_R^T & \mu_X
    \end{array}\right)\,,
\end{equation}
with the $3\times3$ Dirac mass matrix given by $m_D=Y_\nu \langle
\Phi\rangle$. $M_{\mathrm{ISS}}$ being
complex and symmetric, we can use the Takagi factorization to write
\begin{equation}
U^T_\nu M_{\mathrm{ISS}} U_\nu = \text{diag}(m_{n_1},\dots,m_{n_9})\,, \label{Unu}
\end{equation}
where $U_\nu$ is a $9\times 9$ unitary matrix.

A specificity of the ISS model is the presence of a nearly conserved
lepton number. The light neutrino masses are then suppressed by the
small lepton number breaking parameter $\mu_X$ and the heavy
Majorana neutrinos, which have nearly degenerate masses, form
pseudo-Dirac pairs. This can clearly be seen if we consider only one
generation. In the inverse seesaw limit $\mu_X \ll m_D, M_R$, we have
one light neutrino $\nu$ and two heavy neutrinos $N_1\,,N_2$ with
masses
\begin{align}
 m_\nu &\simeq \frac{m_{D}^2}{m_{D}^2+M_{R}^2} \mu_X\,\label{mnu},\\
 m_{N_1,N_2} &\simeq \sqrt{M_{R}^2+m_{D}^2} \mp \frac{M_{R}^2 \mu_X}{2 (m_{D}^2+M_{R}^2)}\,.\label{mN}
\end{align}
With three generations, $M_{\mathrm{ISS}}$ can be diagonalized by
block to give the light neutrino mass matrix, at leading order in the
seesaw expansion parameter $m_D M_R^{-1}$,
\begin{equation}
 \label{MlightLO}
 M_{\mathrm{light}} \simeq m_D M_R^{T-1} \mu_X M_R^{-1} m_D^T\,.
\end{equation}
The next order terms are given in Appendix~\ref{app:NLO}. This
$3\times 3$ complex symmetric mass matrix is diagonalized by
using a unitary matrix identified with the
Pontecorvo-Maki-Nakagawa-Sakata (PMNS) matrix $U_{\rm
  PMNS}$~\cite{pontecorvo1957mesonium,Maki:1962mu}:
\begin{equation} \label{mnulight}
 U_{\rm PMNS}^T M_{\mathrm{light}} U_{\rm PMNS} =
 \mathrm{diag}(m_{n_1}\,, m_{n_2}\,, m_{n_3})\equiv m_\nu\,,
\end{equation}
with $m_{n_1}$, $m_{n_2}$ and $m_{n_3}$ the masses of the three
light neutrinos.

In order to reproduce low-energy neutrino data, different
parameterizations can be introduced. Working in the basis where $M_R$
is diagonal with entries $M_i$, neutrino oscillations are generated by
off-diagonal terms in $m_D$ and $\mu_X$. In a first parameterization,
we can reconstruct $m_D$ as a function of neutrino oscillation data
and high energy parameters. This leads to a Casas-Ibarra
parameterization~\cite{Casas:2001sr} adapted to the inverse seesaw
\begin{equation}
 m_D^T= V^\dagger
 \mathrm{diag}(\sqrt{M_1}\,,\sqrt{M_2}\,,\sqrt{M_3})\; R\;
 \mathrm{diag}(\sqrt{m_{n_1}}\,, \sqrt{m_{n_2}}\,, \sqrt{m_{n_3}})
 U^\dagger_{\rm PMNS}\,,
\label{CasasIbarraISS}
\end{equation}
where $M_1$, $M_2$, $M_3$ are the positive square roots of $M
M^\dagger$ and $M$ is defined by
\begin{equation}
 M=M_R \mu_X^{-1} M_R^T\,.
 \label{CasasIbarraISSM}
\end{equation}
$V$ is a unitary matrix that diagonalize $M$ according to $
M=V^\dagger \mathrm{diag}(M_1\,, M_2\,, M_3) V^*$ and $R$ is a complex
orthogonal matrix that can be expressed as
\begin{equation}
\label{R_Casas}
R = \left( \begin{array}{ccc} c_{2} c_{3} 
& -c_{1} s_{3}-s_1 s_2 c_3& s_{1} s_3- c_1 s_2 c_3\\ c_{2} s_{3} & c_{1} c_{3}-s_{1}s_{2}s_{3} & -s_{1}c_{3}-c_1 s_2 s_3 \\ s_{2}  & s_{1} c_{2} & c_{1}c_{2}\end{array} \right) \,,
\end{equation}
with $c_i= \cos \theta_i$, $s_i = \sin\theta_i$, $\theta_i$ being
arbitrary complex angles.

The other possibility is to use the $\mu_X$-parameterization that was
introduced in ref.~\cite{Arganda:2014dta}, giving
\begin{equation}
\mu_X=M_R^T ~m_D^{-1}~ U_{\rm PMNS}^* m_\nu U_{\rm PMNS}^\dagger~ {m_D^T}^{-1} M_R\,.
\end{equation}
Both parameterizations are based on eq.(\ref{MlightLO}) where only the
leading order term in the seesaw expansion is considered. While this is
sufficient in most of the parameter space, these formulas fail to
reproduce low-energy neutrino data when the active-sterile mixing
becomes very large. Indeed, a large active-sterile mixing corresponds
to a large seesaw expansion parameter $m_D M_R^{-1}$, which makes the
next order terms presented in eq.(\ref{NLOterms}) relevant. Including
the next order terms in the seesaw expansion in the
$\mu_X$-parameterization gives
\begin{align}
  \label{muXparam}
  \begin{split}
    \mu_X \simeq
    & \left(\mathbf{1}-\frac{1}{2} M_R^{*-1} m_D^\dagger m_D M_R^{T-1}
    \right)^{-1} M_R^T m_D^{-1} U_{\rm PMNS}^* m_\nu U_{\rm
      PMNS}^\dagger m_D^{T-1} M_R\, \times\\
    & \left(\mathbf{1}-\frac{1}{2} M_R^{-1} m_D^T m_D^*
      M_R^{\dagger-1}\right)^{-1}\,,
  \end{split}
\end{align}
which allows to better reproduce neutrino oscillation data. The
complete derivation of this formula is given in
Appendix~\ref{app:NLO}.

Finally, we need to specify the couplings between SM particles and the
new fields that are relevant for our calculation of the corrections to
the triple Higgs coupling
$\lambda_{HHH}$. Following ref.~\cite{Ilakovac:1994kj}, we introduce the
$B$ and $C$ matrices defined as
\begin{align}
B_{ij} &= \sum_{k=1}^3 V_{L\,ki}^* U_{\nu\, kj}^{*}\,,\\
C_{ij} &= \sum_{k=1}^3 U_{\nu\, k i} U_{\nu\, kj}^{*}\,,
\end{align}
where $V_{L}$ is the unitary matrix that diagonalizes the charged
lepton mass matrix $M_\mathrm{charged}$ according to
\begin{equation}
 V_L^\dagger\; M_\mathrm{charged}\; V_R = \mathrm{diag}(m_e\,, m_\mu\,, m_\tau)\,,
\end{equation}
with $V_R$ another unitary matrix. In the Feynman-'t Hooft gauge and
in the mass basis, the relevant interaction terms in the Lagrangian
are
\begin{align}
\mathcal{L}_{\rm int}^{Z} &= -\frac{g_2}{4 \cos \theta_W} \bar n_i \slashed{Z} \left[C_{ij} P_L - C_{ij}^* P_R \right] n_j\,,\nonumber\\
\mathcal{L}_{\rm int}^{H} &= -\frac{g_2}{4 m_W} H \bar n_i \left[(C_{ij}  m_{n_i} + C_{ij}^*  m_{n_j}) P_L + (C_{ij}  m_{n_j}+C_{ij}^*  m_{n_i}) P_R \right] n_j\,,\nonumber\\
\mathcal{L}_{\rm int}^{G^0} &= \frac{\imath g_2}{4 m_W} G^0 \bar n_i  \left[- ( C_{ij} m_{n_i} +C_{ij}^* m_{n_j}) P_L + (C_{ij} m_{n_j}+C_{ij}^* m_{n_i}) P_R \right] n_j\,,\nonumber\\
\mathcal{L}_{\rm int}^{W^{\pm}} &= -\frac{g_2}{\sqrt{2}} \bar{l_i} B_{ij} \slashed{W}^{-} P_L n_j + h.c\,,\nonumber\\
\mathcal{L}_{\rm int}^{G^{\pm}} &= \frac{-g_2}{\sqrt{2} m_W} G^{-}\left[\bar{l_i} B_{ij} (m_{l_i} P_L - m_{n_j} P_R) n_j \right] + h.c\,,
\label{eqn:iss-lagrangian}
\end{align}
where $g_2$ is the $\mathrm{SU}(2)$ coupling constant, $\theta_W$ is
the weak mixing angle and $P_L$, $P_R$ are respectively
$(1-\gamma_5)/2$ and $(1+\gamma_5)/2$.

\subsection{Constraints on the ISS model}
\label{sec:constraints}

Strong experimental and theoretical constraints on the parameter space
of the model have to be considered, in particular on the size of the
active-sterile mixing. Our use of the modified Casas-Ibarra or
$\mu_X$-parameterization allows to reproduce neutrino oscillation
data. In our numerical study, we explicitly check the agreement with
the neutrino masses and mixing obtained in the global fit {NuFIT}
3.0~\cite{Esteban:2016qun}. The light neutrino masses are also
chosen to agree with the Planck result~\cite{Ade:2015xua}
\begin{equation}
 \sum_{i=1}^{3} m_{n_i}^{} < 0.23\,\mathrm{eV}\,.
\label{eq:planck}
\end{equation}
The mixing between the active and sterile neutrinos will also induce
deviations from unitarity in the $3\times 3$ sub-matrix $\tilde
U_\mathrm{PMNS}^{}$ of the full $9\times 9$ mixing matrix $U_\nu^{}$,
  that controls the mixing between the light
neutrinos~\cite{Langacker:1988ur,Antusch:2006vwa}. Using a polar
decomposition, this square complex matrix can be expressed as
\begin{equation}
 \tilde U_\mathrm{PMNS}=(I-\eta)\,U_\mathrm{PMNS}\,,
\end{equation}
where $\eta$ is a Hermitian matrix that encodes the deviations from
unitarity. We have included the following constraints from a recent
fit~\cite{Fernandez-Martinez:2016lgt} to electroweak precision
observables, tests of CKM unitarity and tests of lepton universality,
\begin{align}
\label{EWPOconstraints}
 \sqrt{2|\eta_{ee}|}&<0.050\,, 		 & \sqrt{2|\eta_{e\mu}|}<0.026\,, \nonumber \\
 \sqrt{2|\eta_{\mu\mu}|}&<0.021\,, 	 & \sqrt{2|\eta_{e\tau}|}<0.052\,, \nonumber \\
 \sqrt{2|\eta_{\tau\tau}|}&<0.075\,,	 & \sqrt{2|\eta_{\mu\tau}|}<0.035\,.
\end{align}
In the presence of a large active-sterile mixing, the off-diagonal
entries in the neutrino Yukawa couplings $Y_\nu$ might also induce
large branching ratios for lepton flavor violating (LFV) decays. We
have implemented the analytical expressions
from ref.~\cite{Ilakovac:1994kj} for the LFV radiative decays and the LFV
three-body decays. The corresponding experimental upper limits on the
LFV radiative decays~\cite{TheMEG:2016wtm,Aubert:2009ag} are
\begin{align}
 \mathrm{Br}(\mu^+\rightarrow e^+ \gamma)		&< 4.2\times10^{-13}\,, \\
 \mathrm{Br}(\tau^\pm \rightarrow e^\pm \gamma)		&< 3.3\times10^{-8}\,, \\
 \mathrm{Br}(\tau^\pm \rightarrow \mu^\pm \gamma)	&< 4.4\times10^{-8}\,,
\end{align}
at $90\%$~C.L. while the upper limits on LFV three-body
decays~\cite{Bellgardt:1987du,Hayasaka:2010np} are
\begin{align}
 \mathrm{Br}(\mu^+\rightarrow e^+e^+e^-)		&< 1.0\times10^{-12}\,, \\
 \mathrm{Br}(\tau^-\rightarrow e^- e^+ e^- )		&< 2.7\times10^{-8} \,, \\
 \mathrm{Br}(\tau^-\rightarrow \mu^- \mu^+ \mu^-)	&< 2.1\times10^{-8} \,, \\
 \mathrm{Br}(\tau^-\rightarrow e^- \mu^+ \mu^-)		&< 2.7\times10^{-8} \,, \\
 \mathrm{Br}(\tau^-\rightarrow \mu^- e^+ e^-)		&< 1.8\times10^{-8} \,, \\
 \mathrm{Br}(\tau^-\rightarrow e^+ \mu^- \mu^-)		&< 1.7\times10^{-8} \,, \\
 \mathrm{Br}(\tau^-\rightarrow \mu^+ e^- e^-)		&< 1.5\times10^{-8} \,,
\end{align}
at $90\%$~C.L.

We will also require in our study that Yukawa couplings are
perturbative since the complex angles of the R matrix in the
Casas-Ibarra parameterization or the use of $Y_\nu$ as an input
parameter in the $\mu_X$-parameterization can lead to arbitrarily
large entries in $Y_\nu$. We will ensure the perturbativity of the
Yukawa couplings by requiring
\begin{equation}
 \frac{|Y_{ij}|^2}{4 \pi} < 1.5\,,
\end{equation}
for $i,j=1\dots 3$. Since the decay width of heavy neutrinos grows like
$m_n^3$ when $m_n \gg m_H$, we also require that their decay width
verifies, for $i=4\dots 9$,
\begin{equation}
 \Gamma_{n_i}<0.6 m_{n_i}\,,
\label{eq:neutwith}
\end{equation}
in order for the quantum state to be a definite particle. The formulae
used to calculate the heavy neutrino widths are taken from
ref.~\cite{Atre:2009rg}.

%

%
%
\section{Framework of the calculation}
\label{sec:calc}
Our calculation is done in the Feynman-'t Hooft gauge and we use the
Lagrangian of eq.(\ref{eqn:iss-lagrangian}) for the neutrino
interactions. The SM scalar potential is written as 
\begin{align}
V(\Phi) = & -\mu_{}^2 |\Phi|_{}^2 + \lambda |\Phi|_{}^4,
\end{align}
with the Higgs field $\Phi$ given by
\begin{align}
\Phi = & \frac{1}{\sqrt{2}}
         \left(\begin{matrix} \sqrt{2} G^+\\v+H+\imath
                 G^0\end{matrix}\right).
\end{align}
$H$ stands for the Higgs boson, $G_{}^0$ the neutral Goldstone boson, 
$G_{}^\pm$ the charged Goldstone bosons and $v\simeq 246$ GeV is the 
vacuum expectation value (vev) of the Higgs field. We can define the
Higgs tadpole $t_H^{}$, the Higgs mass $M_H^{}$ and the triple Higgs
coupling $\lambda_{HHH}^{}$ as follows,
\begin{align}
  t_H^{} = & -\left\langle \frac{\partial V}{\partial H}
             \right\rangle,\nonumber\\
  M_H^{ 2} = & \phantom{-} \left\langle \frac{\partial_{}^2 V}{\partial H_{}^2}
             \right\rangle,\\
  \lambda_{HHH}^{} = & -\left\langle \frac{\partial_{}^3 V}{\partial
                       H_{}^3} \right\rangle.\nonumber
\end{align}
This helps to redefine the triple Higgs coupling using $t_H^{}$,
$M_H^{}$ and $v$ as input parameters,
\begin{align}
\lambda_{HHH}^{} = 
  & - \frac{3 M_H^2}{v} \left( 1 + \frac{t_H^{}}{v M_H^2}\right).
\end{align}
At tree-level, $t_H^{}=0$ and we recover the usual definition of the
tree-level triple Higgs coupling,
\begin{align}
\lambda^{0} = - \frac{3 M_H^2}{v}.
\end{align}

For the one-loop corrections to the triple Higgs coupling, our set of
input parameters that need to be renormalized in the on-shell (OS)
scheme will be the following:
\begin{align}
M_H^{},\ M_W^{},\ M_Z^{},\ e,\ t_H^{}.
\end{align}
We use the following relations to define the Higgs vev $v$ and the
weak angle $\theta_W^{}$,
\begin{align}
v & = 2\, \frac{M_W^{} \sin\theta_W^{}}{e},\nonumber\\
\sin_{}^2\theta_W^{} & =  1-\frac{M_W^2}{M_Z^2},
\end{align}
as well as $e^2_{}= 4\pi\alpha$.

We require that we have no tadpoles at one loop:
\begin{equation}
t^{(1)}_H + \delta t_H^{} = 0 \Rightarrow \delta t_H^{} = - t^{(1)}_H,
\end{equation}
with $t^{(1)}_H$ being the one-loop un-renormalized contributions to
$t_H^{}$. For the other parameters we introduce their counter-terms as 
follows,
\begin{align}
 & M_H^2 \rightarrow M_H^2 + \delta M_H^2,\nonumber\\
 & M_W^2 \rightarrow M_W^2 + \delta M_W^2,\nonumber\\
 & M_Z^2 \rightarrow M_Z^2 + \delta M_Z^2,\\
 & e \rightarrow (1+\delta Z_e^{})\, e,\nonumber\\
 & H \rightarrow \sqrt{Z_H^{}} H = \left(1+ \frac12 \delta
   Z_H^{}\right) H.\nonumber
\end{align}

The full renormalized one-loop triple Higgs coupling is finally
\begin{equation}
\lambda^{1r}_{HHH}(q_H^*) = \lambda^0 + \lambda_{HHH}^{(1)}(q_H^*) +
\delta\lambda_{HHH}^{}\,,
\end{equation}
with
\begin{align}
\delta\lambda_{HHH}^{}  = \lambda_{}^0 
  & \left[\frac32 \delta Z_H^{} + \delta t_H^{} \frac{e}{2 M_W^{}
    \sin\theta_W^{} M_H^2} +\delta Z_e^{} + \frac{\delta M_H^2}{M_H^2}
    - \frac{\delta M_W^2}{2 M_W^2} \right.\nonumber\\
  & \left. + \frac12 \frac{\cos_{}^2\theta_W^{}}{\sin_{}^2\theta_W^{}}
    \left( \frac{\delta M_W^2}{M_W^2}-\frac{\delta
    M_Z^2}{M_Z^2}\right) \right],
\end{align}
and $\lambda^{(1)}_{HHH}(q_H^*)$ stands for the un-renormalized
one-loop contributions to the process $H^*_{}\to H H$ with the
momentum $q_H^*$ for the off-shell Higgs boson $H^*_{}$. For the
numerical analysis carried in the next section, we define the
deviation induced by the BSM contribution $\Delta^{\rm BSM}_{}$ as
\begin{align}
\Delta^{\rm BSM}_{} = \frac{1}{\lambda^{1r, {\rm
  SM}}_{HHH}} \left(\lambda^{1r}_{HHH} - \lambda^{1r,
  {\rm SM}}_{HHH} \right),
\end{align}
where $\lambda^{1r, {\rm SM}}_{HHH}$ stands for the renormalized one-loop SM
contribution without the light neutrinos.

Introducing the notation $\Sigma_{XY}^{}$ for the self-energy of the
process $X\to Y$, we use the usual OS conditions for $M_W^{}$,
$M_Z^{}$ and $M_H^{}$,
\begin{align}
\delta M_W^2 & = \re \Sigma^T_{WW}(M_H^2),\nonumber\\
 \delta M_Z^2 & = \re \Sigma^T_{ZZ}(M_H^2),\nonumber\\
\delta M_H^2 & = \re \Sigma_{HH}^{}(M_H^2).
\end{align}
 For the electric charge $e$ we use the following condition to be
 independent from the light fermion
 masses~\cite{Denner:1991kt,Nhung:2013lpa},
\begin{align}
\delta Z_e^{}  =
  \frac{\sin\theta_W^{}}{\cos\theta_W^{}}\frac{\re\Sigma^T_{\gamma
  Z}(0)}{M_Z^2} - \frac{\re\Sigma^T_{\gamma \gamma}(M_Z^2)}{M_Z^2}.
\end{align}
For the Higgs field renormalization we have
\begin{align}
\delta Z_H^{} & = -\re \left.\frac{\partial \Sigma_{HH}^{}
                (k_{}^2)}{\partial k_{}^2}\right|^{}_{k_{}^2=M_H^2}.
\end{align}

\begin{figure}[t]
  \bec
  \includegraphics[width=\textwidth,clip]{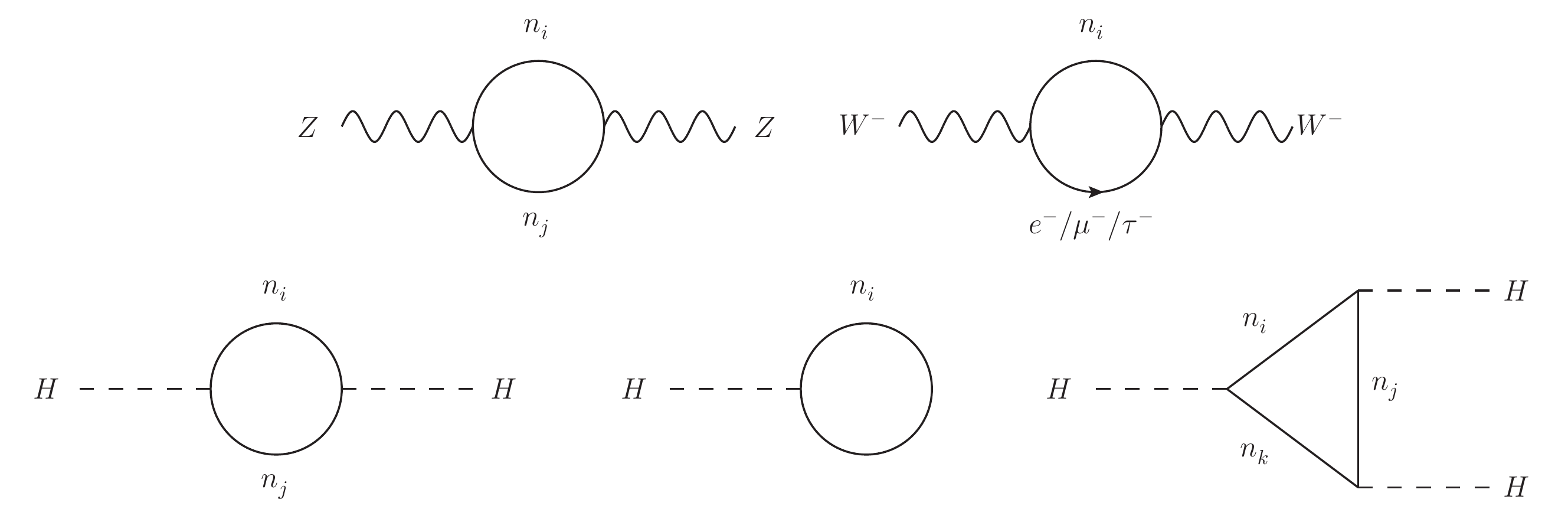}
  \caption{Feynman diagrams for the neutrino contributions to the
    one-loop $W$ and $Z$ boson self-energies (upper line) and the
    one-loop Higgs boson self energy, tadpole and triple Higgs
    coupling (lower line). In all diagrams, the indices i/j/k run from
    1 to 9.}
  \label{fig:feyndiag}
  \eec
\end{figure}

The neutrino interactions induce changes in the $W$ and $Z$
self-energies as well as in the Higgs tadpole, self-energy and
self-couplings. We display in Fig.~\ref{fig:feyndiag} the Feynman
diagrams for the neutrino contributions to the $W$, $Z$ and Higgs
bosons self-energies, the Higgs tadpole and the one-loop
un-renormalized triple Higgs coupling. We also collect in
Appendix~\ref{app:HHH} the analytical expressions of the neutrino
contributions to $\delta M_W^{}$, $\delta M_Z^{}$, $\delta t_H^{}$,
$\Sigma_{HH}^{}$ and $\lambda_{HHH}^{(1)}$. They were obtained using
{\tt FeynArts 2.7}~\cite{Hahn:2000kx} and {\tt FormCalc
  7.5}~\cite{Hahn:1998yk}, in which we have implemented our own Model
File for the ISS model. The scalar and tensor loop
functions~\cite{'tHooft:1978xw,Passarino:1978jh} have been evaluated with {\tt LoopTools
  2.13}~\cite{vanOldenborgh:1990yc,Hahn:1998yk,Hahn:2006qw}. We have
checked numerically that the UV divergences cancel in the final result
and that the renormalized one-loop triple Higgs coupling does not
depend on the choice of the renormalization scale.

%
%
\section{Numerical results}
\label{sec:pheno}

We present in this section the phenomenological study of the one-loop
corrected triple Higgs coupling and the dependence of the corrections
induced by the heavy neutrinos on the relevant input parameters of the
ISS model. The SM parameters are taken from the Particle Data Group
(PDG)~\cite{Olive:2016xmw} (with the exception of the SM Higgs boson
mass) and read as
\begin{eqnarray}
  m_t^{\rm pole} = 173.5~\text{GeV},\ \ & m_b^{\rm pole} =
  4.77~\text{GeV},\ \ & m_c^{\rm pole} = 1.42~\text{GeV},\nonumber\\
  M_W^{} = 80.385~\text{GeV},\ \ & M_Z^{} = 91.1876~\text{GeV},\ \
  & M_H^{} = 125~\text{GeV},\\
  m_e^{} = 0.511~\text{MeV},\ \ & m_\mu^{} = 105.7~\text{MeV},\ \
  & m_\tau^{} = 1.777~\text{GeV},\nonumber\\
  \phantom{m_e^{} = 0.511~\text{MeV},\ \ } 
   & \alpha^{-1}_{}(M_Z^2) = 127.934.\ \ &
    \phantom{m_\tau^{} = 1.777~\text{GeV},} \nonumber
\end{eqnarray}

The up-, down- and strange-quark masses are also taken from the PDG,
but their impact on the calculation is negligible so that we do not
list them here. The lightest neutrino mass is chosen as
\begin{align}
m_{n_1}^{} = 0.01~\text{eV},
\end{align}
to comply with cosmological constraints as stated in
eq.(\ref{eq:planck}). We have explicitly checked that choosing a
smaller mass for $n_1$ does not qualitatively modify our results and
would only induce negligible numerical corrections to our final
conclusions. We chose the normal ordering for the neutrino
masses and the light neutrino mixing parameters are taken from NuFIT
3.0~\cite{Esteban:2016qun}, with $\delta_\mathrm{CP}=0$. Since the
contributions of the light neutrinos are negligible and flavor
constraints do not play an important role in our final conclusion, we
do not expect our conclusion to change if we consider the inverted
ordering. 

In our study, we will focus on two choices for the off-shell Higgs
momentum $q^*_H=500$~GeV and $q^*_H=2500$~GeV. These choices follow
from the behavior of the BSM corrections that exhibit a similar
dependence on $q^*_H$ between the ISS model and the simplified Dirac
3+1 model that was studied in~ref.\cite{Baglio:2016ijw}. In
particular, the maximal negative deviation was obtained for
$q^*_H=500$~GeV while the maximal positive deviation was obtained for
large off-shell Higgs momenta. To facilitate the comparison between
the Majorana ISS case and the simplified Dirac case we take the same
fixed values of  $q^*_H$ as in~ref.\cite{Baglio:2016ijw} in all the
scans.

\subsection{Casas-Ibarra parameterization}

In order to get an insight into the parameter space of the ISS model
we perform a scan in a Casas-Ibarra parameterization, see
eq.(\ref{CasasIbarraISS}). The goal is to get an idea of the
corrections that are obtained in this parameterization and the impact
of the constraints on the model. We perform a random scan using a flat
prior on the three real rotation angles $\theta_{1/2/3}^{}$ of the
orthogonal matrix $R$ and a logarithmic prior on both  the lepton
number violating term $\mu_X^{}$ so that the Majorana mass term is
$\mu_X^{}\equiv \mu_X^{}  {\rm I}_3^{}$, and the mass term $M_R^{}$
so that the matrix $M_R^{}$ is $M_R^{} \equiv M_R^{} {\rm I}_3^{}$. 
We take all mass and rotation matrices to be real
in order to avoid generating CP violation.
We use 180\! 000 randomly generated points in the following parameter
range,
\begin{eqnarray}
  0 & \leq \theta_i^{} & \leq 2\pi, \ (i=1\dots 3),\nonumber\\ 
  0.2~\text{TeV} & \leq M_R^{} & \leq 1000~\text{TeV}, \\
  7.00\times 10^{-4}_{}~\text{eV} & \leq \mu_X^{} & \leq 8.26\times
  10^{4}~\text{eV}. \nonumber
\end{eqnarray}
The range choice for the parameter $\mu_X^{}$ follows 
$\displaystyle\mu_X^{\rm min[max]} = \left(M_R^{\rm min[max]}\right)^2_{}
\frac{m_{n_1^{}}^{}}{3\pi v^2_{} [2 v^2_{}]}$, see
eq.(\ref{mnu}). Heavy neutrino masses below 200 GeV are better
probed with direct searches at
colliders~\cite{Antusch:2006vwa} (see also ref.~\cite{Antusch:2016ejd}
and references therein), thus we do not take $M_R^{} < 0.2$~TeV.

\begin{figure}[t!]
\centering
\hspace{10mm}\includegraphics[scale=0.85,clip]{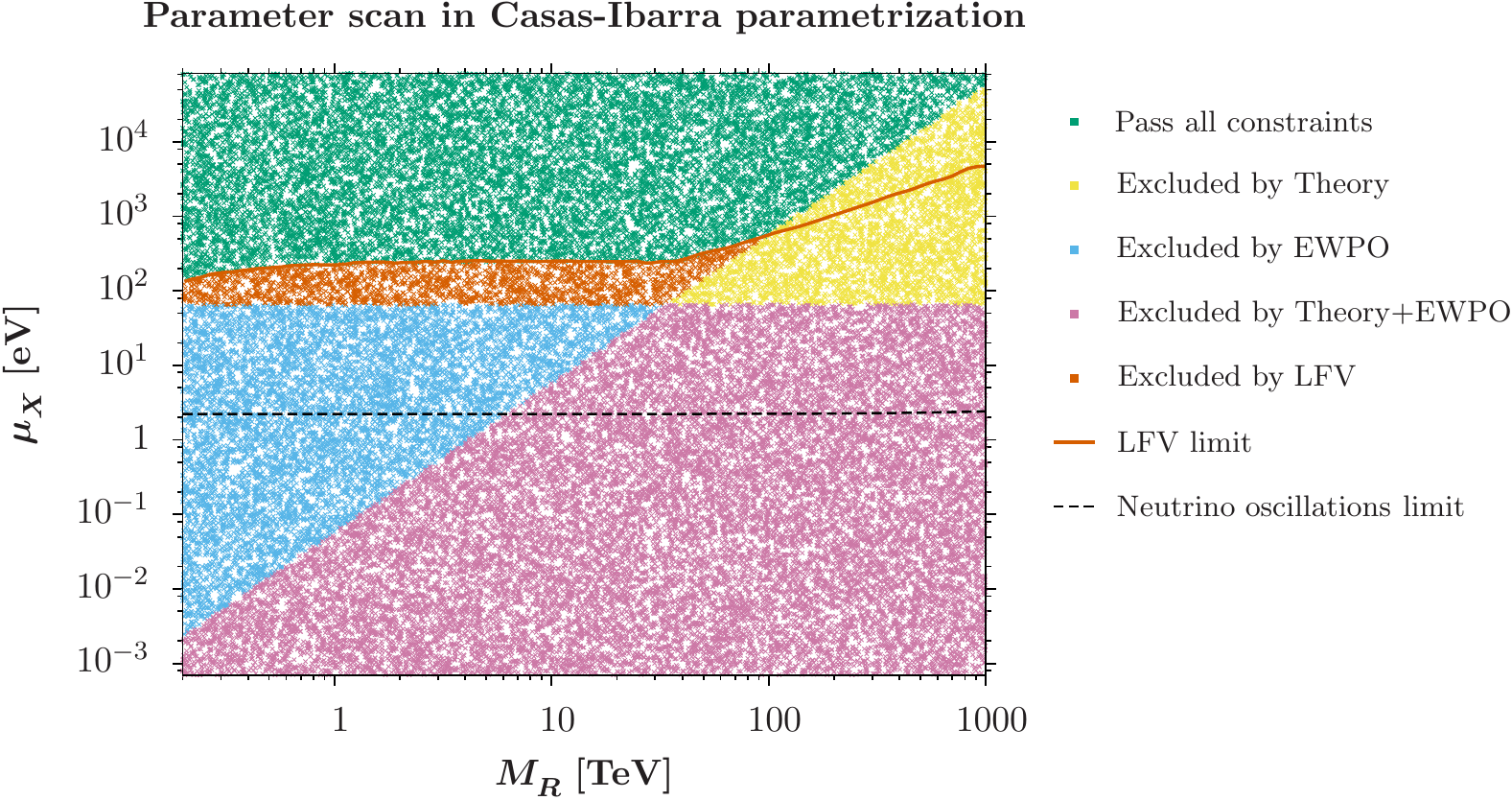}\bigskip

\includegraphics[scale=0.85,clip]{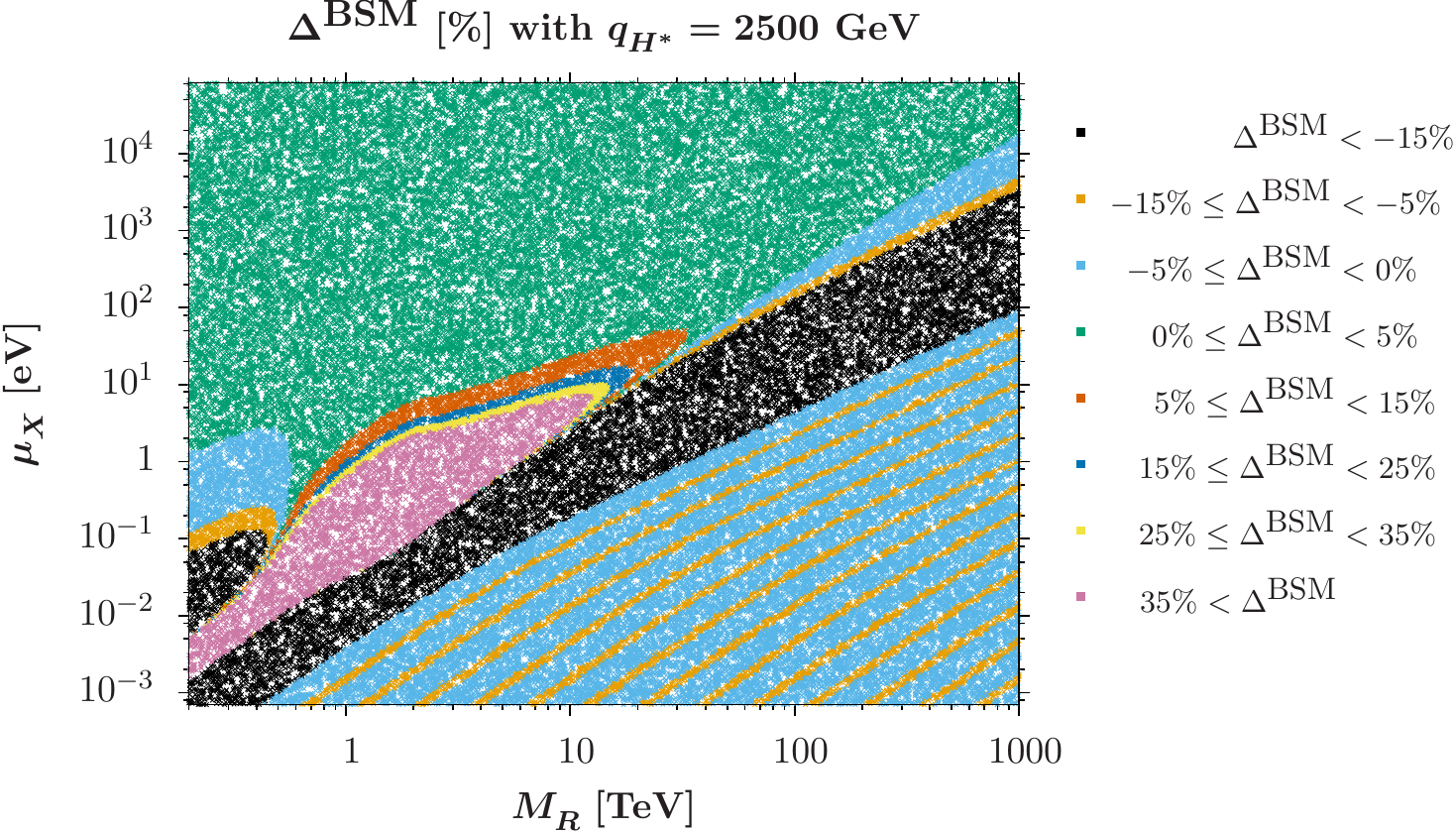}
 \caption{Random scan of the parameter space with 180\! 000 points in
   the Casas-Ibarra parameterization as a function of $M_R^{}$ (in
   TeV) and of $\mu_X^{}$ (in eV). Upper row: Map of the points according
   to the constraints on the model. The vermilion (solid) line stands
   for the LFV constraints and the black (dashed) line stands for the
   constraints coming from neutrino oscillations. All points below
   these lines are excluded. In green, the points that pass all
   the constraints; in yellow, the points that are excluded by
   theory constraints; in blue, the points that are excluded by
   EWPO; in purple, the points that are excluded both by EWPO and theory
   constraints. Lower row: Map of $\Delta^{\rm BSM}$ correction (in
   percent). In black: $\Delta^{\rm BSM}< -15\%$; in orange: $-15\%
   \leq \Delta^{\rm BSM}< -5\%$; in light blue: $-5\% \leq \Delta^{\rm
     BSM}< 0\%$; in green: $0\% \leq \Delta^{\rm BSM}< 5\%$; in
   vermilion: $5\% \leq \Delta^{\rm BSM}< 15\%$; in blue: $15\% \leq
   \Delta^{\rm BSM}< 25\%$; in yellow: $25\% \leq \Delta^{\rm BSM}<
   35\%$; in purple: $\Delta^{\rm BSM}> 35\%$.}
 \label{fig:casas-ibarra}
 \end{figure}

The result of our scan is displayed in Fig.~\ref{fig:casas-ibarra}
(upper row) in the $M_R^{}-\mu_X^{}$ plane. The top-right corner (in
yellow) of the parameter space is excluded by theory constraints,
essentially the perturbativity of the neutrino Yukawa couplings. The
region in light blue is excluded by EWPO, while the region in purple
is excluded both by EWPO and theory constraints. The dashed black line
displays the limit coming from neutrino oscillations. This comes from
a breakdown of the leading-order Casas-Ibarra parameterization when the
active-sterile mixing is too large, as is evidenced by the flat
behavior in $M_R$. In variants of the type I seesaw including the
inverse seesaw, the active-sterile mixing is proportional to the
seesaw expansion parameter $m_D M_R^{-1}$. However, in the
Casas-Ibarra parametrization, $m_D$ grows linearly with $M_R$, see
eqs.(\ref{CasasIbarraISS})-(\ref{CasasIbarraISSM}). As a consequence,
$m_D M_R^{-1}$ appears constant in $M_R$ but increases when $\mu_X$
decreases according to
\begin{equation}
 \frac{m_D}{M_R}\sim \sqrt{\frac{m_\nu}{\mu_X}}\,.
 \label{scalingCI}
\end{equation}
The breakdown happens for $\mu_X\sim3$~eV, which in turns roughly
corresponds to $m_D/M_R\sim0.1$ when taking $m_\nu=m_{n_3}$. It is
worth noting that this value can be predicted from
eq.(\ref{NLOterms}), where next-order corrections to the light
neutrino mass matrix appear at $\mathcal{O}(m_D^2/M_R^2)$, and from
the current error on $\Delta m^2$ being at the percent
level~\cite{Esteban:2016qun}. The most stringent experimental
constraint comes from LFV observables as displayed by the solid
vermilion line. The top-left corner (in green) is allowed by all the
constraints.

This scan has to be compared to the map of $\Delta^{\rm BSM}_{}$
displayed in Fig.~\ref{fig:casas-ibarra} (lower row), fixing the
off-shell Higgs momentum at $q^*_H=2500$~GeV. The parameter space
passing all the constraints only contains corrections up to $\sim
+1\%$. The most interesting regions are in vermilion, blue, yellow
and purple where $\Delta^{\rm BSM}_{}$ reaches $+15\%$, $+25\%$,
$+35\%$ and more than $+35\%$, respectively. In order to enter these
regions, it is needed to escape the LFV constraints as much as
possible. Following ref.~\cite{Arganda:2014dta} we will investigate
this region using the $\mu_X^{}$-parameterization and start with the
case of degenerate heavy neutrinos.

\subsection{Degenerate heavy neutrinos}

The scan in the Casas-Ibarra parameterization displayed in
Fig.~\ref{fig:casas-ibarra} shows that the most stringent constraints
come from LFV observables. In order to maximize the effects on the
triple Higgs coupling we want to escape these constraints and we
require for example $(Y_\nu^{} Y_{\nu}^\dagger)_{1 2}^{} = 0$ since
decays that involve a $\mu-e$ transition usually give the strongest
constraints. This leads to either a diagonal Yukawa matrix or a Yukawa
texture as defined in ref.~\cite{Arganda:2014dta}, with degenerate
heavy neutrinos, $M_R^{} \propto {\rm I}_3^{}$.

\begin{figure}[t]
 \bec
 \begin{minipage}{0.50\textwidth}
   \includegraphics[width=0.99\textwidth,clip]{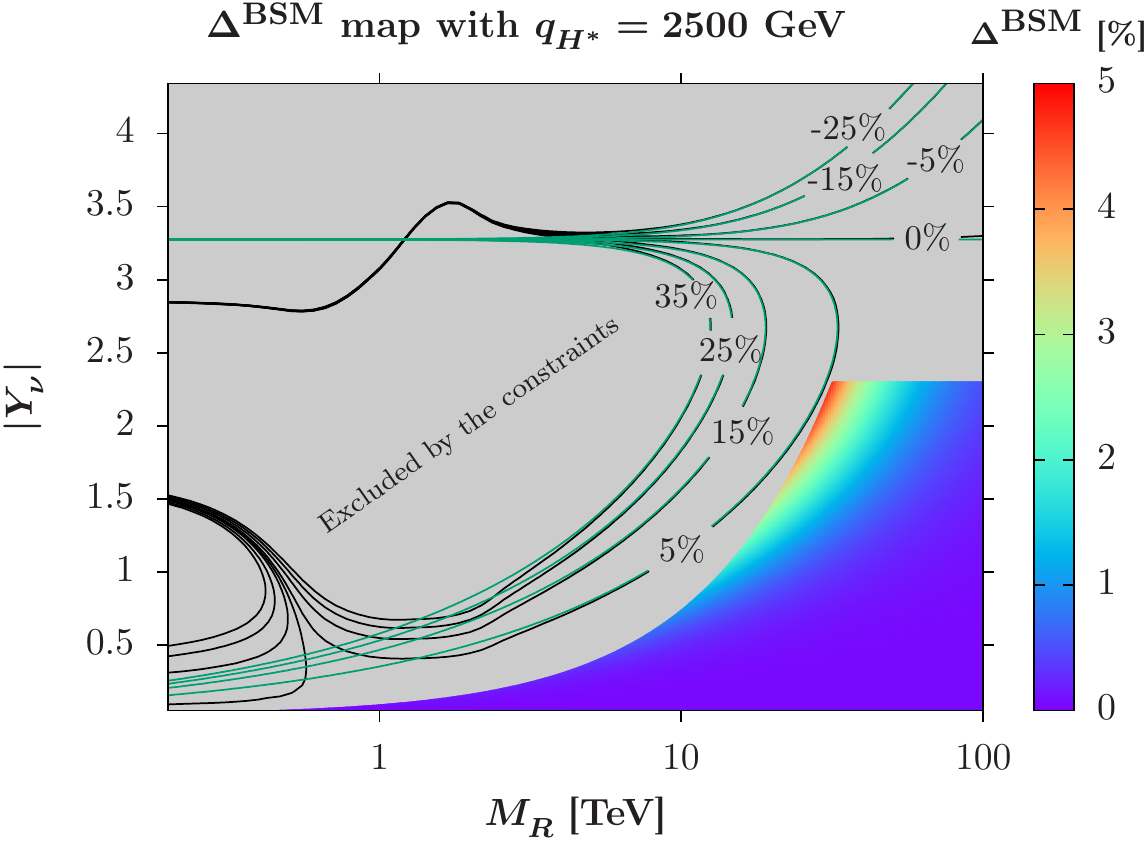}
 \end{minipage}\hspace{1mm}
 \begin{minipage}{0.48\textwidth}
   \includegraphics[width=0.9\textwidth,clip]{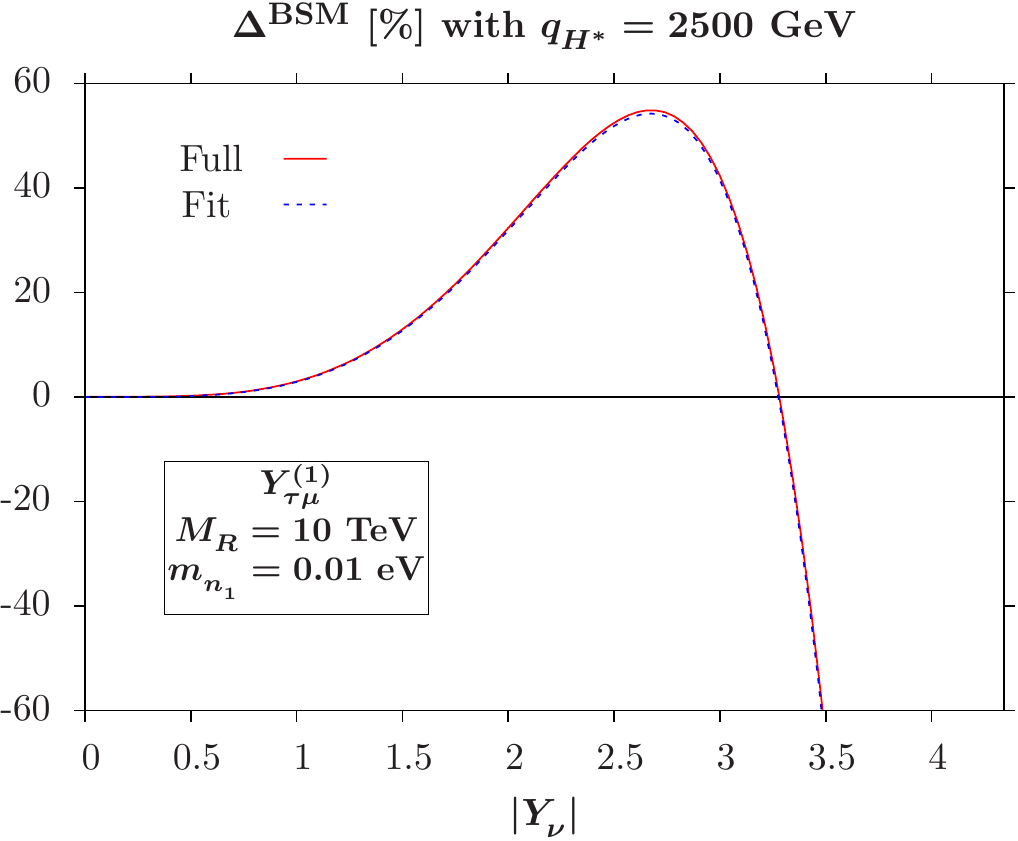}
 \end{minipage}
 \caption{Left: Contour map of the heavy neutrino correction
   $\Delta^{\rm BSM}_{}$ to the triple Higgs coupling
   $\lambda_{HHH}^{}$ (in percent) as a function of the neutrino
   parameters $M_R^{}$ (in TeV) and $|Y_\nu^{}|$ in the
   $\mu_X^{}$-parameterization. The Yukawa texture $Y_{\tau\mu}^{(1)}$
   defined in eq.(\ref{eq:texture}) is used and the off-shell Higgs
   boson momentum is fixed to $q^*_H = 2500$ GeV. The gray area is
   excluded by the constraints on the model. The green lines are
   the approximated contour lines using eq.(\ref{eq:approx}) while the
   black lines correspond to the full calculation. Right: The heavy
   neutrino correction $\Delta^{\rm BSM}_{}$ (in percent) as a
   function of the Yukawa scaling parameter $|Y_\nu^{}|$, in the
   $\mu_X^{}$-parameterization with the texture $Y_{\tau\mu}^{(1)}$. We
   have fixed the other input parameters for the neutrino sector as
   $M_R^{} = 10$~TeV and $m_{n_1^{}}=0.01$~eV. The red (solid) curve
   corresponds to the full calculation, the blue (dashed) curve to the
   approximate result obtained with eq.(\ref{eq:approx}).}
 \label{fig:texture}
 \eec
 \end{figure}

We investigate in this sub-section the case of the degenerate heavy
neutrinos in a \linebreak\mbox{$\mu_X^{}$-parameterization} with the
texture $Y_{\tau\mu}^{(1)}$ taken from ref.~\cite{Arganda:2014dta} and
defined below,
\begin{align}
  Y^{(1)}_{\tau\mu} = |Y_\nu^{}| \left(
  \begin{matrix}
    0 & 1 & -1\\
    0.9 & 1 & 1\\
    1 & 1 & 1
  \end{matrix}\right).
 \label{eq:texture}
\end{align}
We display in Fig.~\ref{fig:texture} (left) the two-dimensional scan
in the plane $(M_R^{},|Y_\nu^{}|)$ where $M_R^{}$ represents the
common scaling factor of the $3\times 3$ diagonal mass matrix
$M_R^{}$. The off-shell Higgs momentum is again fixed at
$q^*_H=2500$~GeV. A large part of the parameter space is excluded and
a maximum of $\Delta^{\rm BSM} \sim +5\%$ can be reached at $M_R
\simeq 13$~TeV. When compared to the Casas-Ibarra scan, this is the
expected order of magnitude for the correction when entering the
vermilion region which is excluded by LFV observables only. For large
$M_R^{}$ the most important constraint is the neutrino width
(\ref{eq:neutwith}). For lower $M_R^{}$ the constraints are driven by
the violation of the unitarity of the $3\times 3$ matrix
$\tilde{U}_{\rm PMNS}^{}$ controlling the mixing between the light
neutrinos.

To get an insight into the behavior of the contour lines in
Fig.~\ref{fig:texture} (left) we display a one-dimensional plot of the
neutrino correction $\Delta^{\rm BSM}_{}$ at a given $M_R^{}=10$~TeV,
as a function of the Yukawa scaling factor $|Y_\nu^{}|$, in
Fig.~\ref{fig:texture}  (right). The correction is negligible for low
Yukawa scaling factors, then rises to a maximum at $\sim +60\%$ at
$|Y_{\nu}^{}| \simeq 2.5$ before dropping rapidly and eventually
becoming negative for large Yukawa scaling factors.

From this behavior we devise the following approximate formula to
reproduce $\Delta^{\rm BSM}$ at $M_R^{} > 3$~TeV,

 \begin{align}
 \Delta^{\rm BSM}_{\rm approx} = \frac{(1~\text{TeV})_{}^2}{M_R^2} \left( 8.45\, {\rm Tr}
   (Y_\nu^{} Y_\nu^\dagger Y_\nu^{} Y_\nu^\dagger) - 0.145\, {\rm Tr}
   (Y_\nu^{}   Y_\nu^\dagger Y_\nu^{} Y_\nu^\dagger Y_\nu^{}
   Y_\nu^\dagger)\right).
 \label{eq:approx}
 \end{align}

The numerical coefficients are found to be universal in term of the
parameters of the model and only depend on the kinematics of the
off-shell Higgs boson, for the case of the three textures of
ref.~\cite{Arganda:2014dta} as well as for the case of a diagonal
texture. The dependence of the numerical coefficients on the
  kinematics of the off-shell Higgs boson is expected, as when
  compared to the full calculation they would result from the loop
  functions depending on $q^*_H$, see Appendix~\ref{app:HHH}. It is
expected that eq.(\ref{eq:approx}) be valid for the whole class of
textures introduced in ref.~\cite{Arganda:2014dta}. At a given 
$M_R^{}>3$~TeV, the approximate formula in eq.(\ref{eq:approx}) is
driven at low $|Y_{\nu}^{}|$ by the positive contribution and by the
negative contribution at high $|Y_{\nu}^{}|$, the latter falling more
rapidly than the positive increase at low $|Y_{\nu}^{}|$. This
reproduces the behavior seen in Fig.~\ref{fig:texture} (right) where
the result of the fit is also displayed. We can also reproduce the
contour lines for high $M_R^{}$ in Fig.~\ref{fig:texture} (left) as
seen from the  green contour lines coming from the fit, that agree
to a very  good extent with the full contour lines for $M_R> 3$~TeV.

The approximate formula in eq.(\ref{eq:approx}) implies that the best
way to maximize the neutrino effects on the triple Higgs coupling
would be to maximize the ratio $\displaystyle \frac{{\rm Tr} (Y_\nu^{}
  Y_\nu^\dagger Y_\nu^{}  Y_\nu^\dagger)}{{\rm Tr} (Y_\nu^{}
  Y_\nu^\dagger Y_\nu^{} Y_\nu^\dagger Y_\nu^{} Y_\nu^\dagger)}$. The
Yukawa couplings being real and limited by perturbativity
requirements, this leads to the choice of a diagonal texture,
$Y_\nu^{} \propto {\rm I}_3^{}$. This will be considered in the next
sub-section, but with the condition of degenerate heavy neutrinos
being relaxed. In such a way the constraints on the non-unitarity of
the matrix $\tilde{U}_{\rm PMNS}^{}$ are softened and the blue region
of the Casas-Ibarra scan of Fig.~\ref{fig:casas-ibarra}, excluded by
EWPO as well as by LFV observables, moves down.

\subsection{Hierarchical heavy neutrinos}

The analysis carried in the previous sub-section has lead us to
consider a diagonal Yukawa matrix, $Y_\nu^{} = |Y_\nu^{}| {\rm
  I}_3^{}$. In order to reduce as much as possible the impact of
unitarity constraints on $\eta$ for the matrix $\tilde{U}_{\rm
  PMNS}^{}$, we chose hierarchical heavy neutrinos with $M_R^{} = {\rm
  diag}(M_{R^{}_1}^{},M_{R^{}_2}^{},M_{R^{}_3}^{})$ and we still work
in the $\mu_X^{}$-parameterization. More specifically, for illustrative
purpose within this class of parameters, we chose
\begin{align}
M_{R^{}_1}^{} = 1.51 M_R^{},\ \ M_{R^{}_2}^{} = 3.59 M_R^{},\ \
  M_{R^{}_3}^{} = M_R^{}, 
\label{eq:hierarchical}
\end{align}
with $M_R^{}$ being a rescaling factor that is varied between 200 GeV
and 20 TeV. This ensures that all the diagonal constraints of
eq.(\ref{EWPOconstraints}) have the same impact on our study. This
specific choice maximizes the individual contribution of each heavy
neutrino, which in turns will maximize the one-loop correction to the
triple Higgs coupling originating from the leptonic sector. Other
choices for the heavy neutrino masses will only reduce the allowed
maximum value of the triple Higgs coupling deviation from the SM.

\begin{figure}[t]
 \bec
 \begin{minipage}{0.49\textwidth}
   \includegraphics[width=0.99\textwidth,clip]{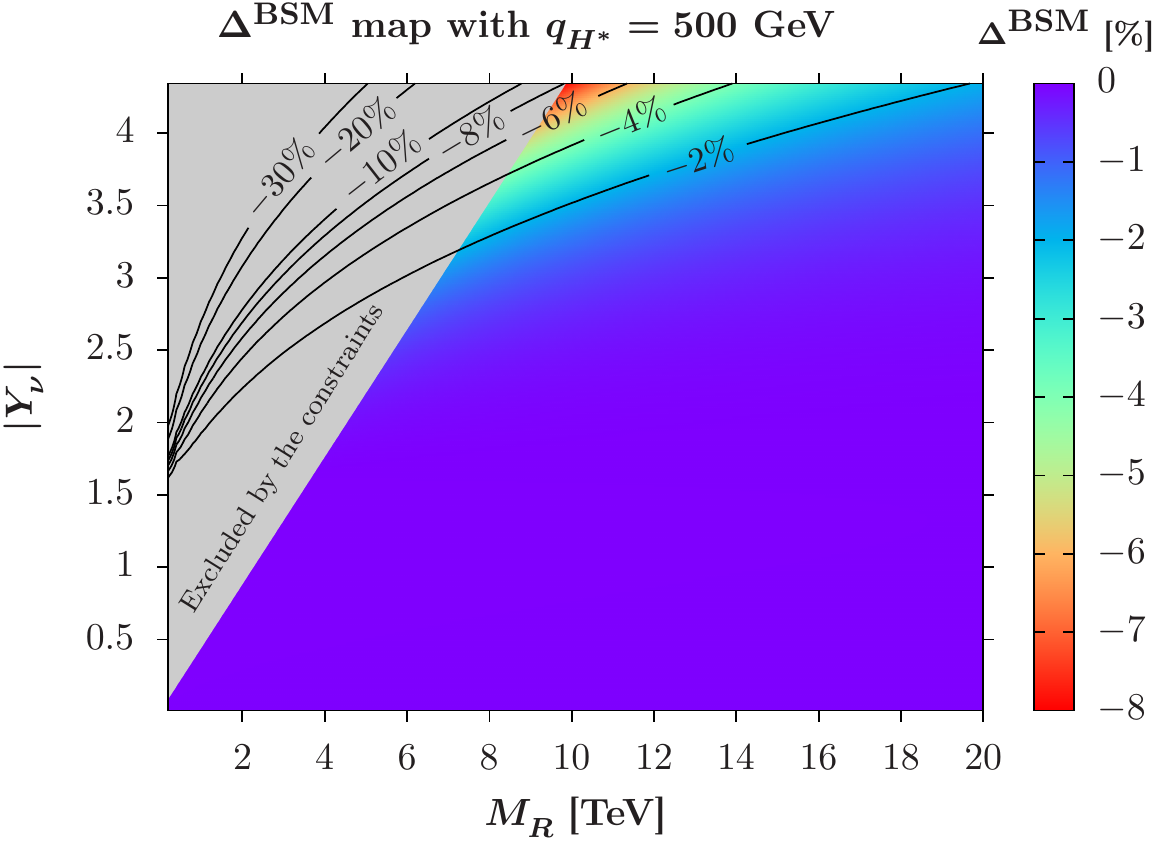}
 \end{minipage}\hspace{1mm}
 \begin{minipage}{0.49\textwidth}
   \includegraphics[width=0.99\textwidth,clip]{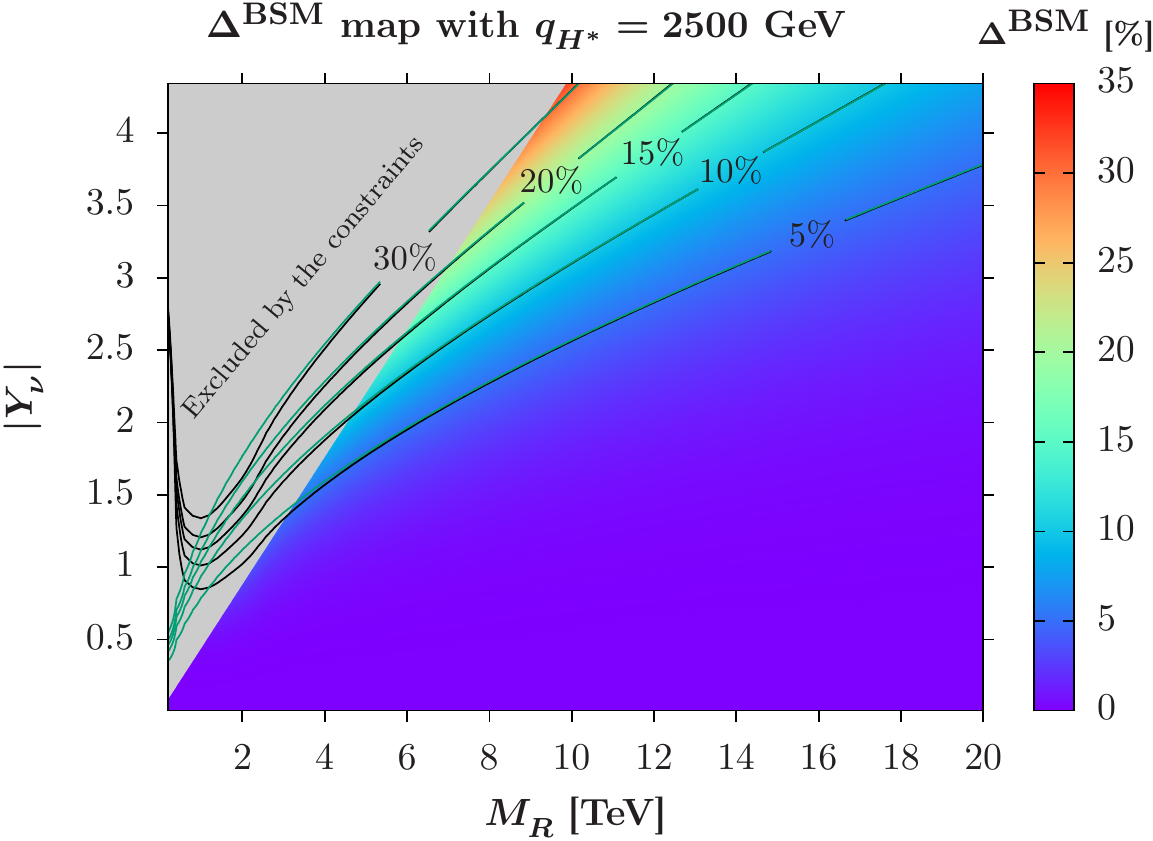}
 \end{minipage}
 \caption{Contour map of the heavy neutrino correction
   $\Delta^{\rm BSM}_{}$ to the triple Higgs coupling
   $\lambda_{HHH}^{}$ (in percent) as a function of the neutrino
   parameters $M_R^{}$ (in TeV) and $|Y_\nu^{}|$ in the
   $\mu_X^{}$-parameterization, using a diagonal Yukawa texture and a
   hierarchical heavy neutrino mass matrix with the parameters defined
   in eq.(\ref{eq:hierarchical}). The off-shell Higgs boson momentum
   is fixed to $q^*_H = 500$~GeV (left) and $q^*_H = 2500$~GeV
   (right). The gray area is excluded by the constraints on the model
   and the green lines on the right figure are the approximated
   contour lines using eq.(\ref{eq:approx}) with a common rescaling
   factor $0.51$, while the black lines correspond to the full
   calculation.}
 \label{fig:hierarchical}
 \eec
 \end{figure}

The result of the parameter scan in the $M_R^{}-|Y_\nu^{}|$ plane is
displayed in Fig.~\ref{fig:hierarchical}. On the left-hand side, we
display the map of $\Delta^{\rm BSM}_{}$ for an off-shell Higgs
momentum $q_H^* = 500$~GeV. As already expected by the analysis in the
simplified model of ref.~\cite{Baglio:2016ijw}, the heavy neutrino
corrections are negative, and they reach a minimum of $\sim -8\%$,
close to the minimum that was obtained in the simplified model. The
most interesting results are displayed in the right-hand side of
Fig.~\ref{fig:hierarchical}, for $q_H^*=2500$ GeV. The corrections can
now reach a maximum of $\sim +30\%$, similar to what has been obtained
in the case of a simplified model. The corrections are generically
bigger in the ISS model than in the simplified model, but the
constraints are also stronger, reducing the heavy neutrino corrections
back to the maximum obtained in the simplified model. This also
confirms in a realistic, renormalizable, low-scale seesaw model that
heavy Majorana neutrinos can induce sizable deviation to the triple
Higgs coupling.

As a further test of our approximate formula for the heavy neutrino
corrections, the green lines in Fig.~\ref{fig:hierarchical} are the
approximate contour lines obtained using eq.(\ref{eq:approx}) but
rescaled with a common factor $\gamma=0.51$. This type of rescaling
was expected as now the heavy neutrino mass matrix is not
proportional to the identity matrix anymore. Once again we obtain a very good
approximation for $M_R^{} > 3$~TeV, and in particular in the
region allowed by the constraints. This approximate formula thus
describes well the behavior of $\Delta^{\rm BSM}_{}$ in the allowed
region of the parameter space, for $q^*_H =2500$~GeV.

\begin{figure}[t]
   \bec 
   \includegraphics[scale=1.0,clip]{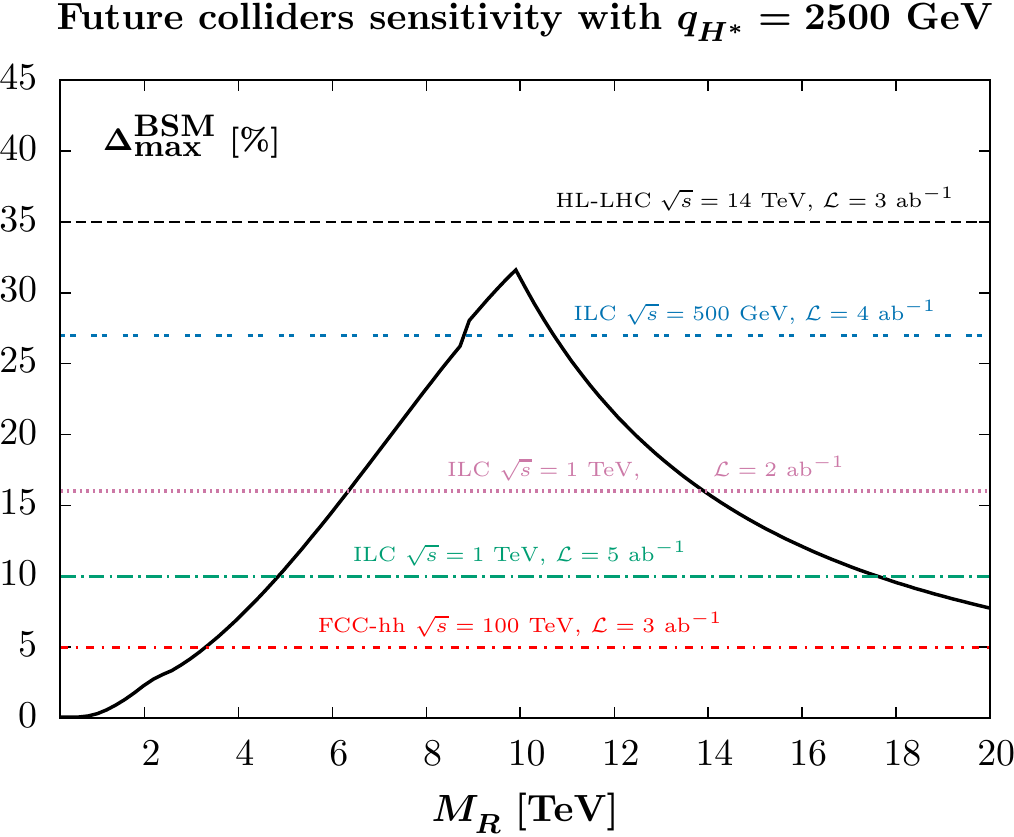}
   \caption{The maximally allowed deviation $\Delta^{\rm BSM}_{\rm
       max}$ (in percent) as a function of the heavy neutrino mass
     parameter $M_R^{}$ (in TeV), compared to the currently expected
     sensitivities for the HL-LHC and the future ILC (with different
     integrated luminosities and center-of-mass energies $\sqrt{s}$)
     and FCC-hh colliders. The solid black line displays $\Delta^{\rm
       BSM}_{\rm max}$, the dashed black line is the LHC-LHC
     sensitivity at 3 ab$^{-1}_{}$, the double-dotted blue line is the
     ILC sensitivity at 4 ab$^{-1}_{}$ with $\sqrt{s}=500$ GeV, the
     dotted line is the ILC sensitivity at 2 ab$^{-1}_{}$ with
     $\sqrt{s}=1$ TeV, the green long dash-dotted line is the ILC
     sensitivity at 5 ab$^{-1}_{}$ with $\sqrt{s}=1$ TeV, and the red
     dash-dotted line is the FCC-hh sensitivity at 3 ab$^{-1}_{}$.}
   \label{fig:sensitivity}
   \eec
 \end{figure}

We end this section with a comparison with the currently expected
sensitivity to the triple Higgs coupling at the HL-LHC and at the
future planned colliders. The sensitivities to the SM triple Higgs
coupling are defined by its measure extracted from the Higgs pair
production yields. As stated for example in
refs.~\cite{Baglio:2012np,Frederix:2014hta}, a precision of $\sim
50\%$ on the total cross section leads to a precision of $\sim 50\%$
on the SM triple Higgs coupling. The sensitivity for the HL-LHC
follows from ref.~\cite{CMS-PAS-FTR-15-002} (see also
ref.~\cite{Campana:2016cqm}), scaled by a factor of 
$1/\sqrt{2}$ to account for both ATLAS and CMS accumulated data, while
the sensitivity for the future colliders follow from
refs.~\cite{Fujii:2015jha,He:2015spf}. For the FCC-hh we do the same
as for the HL-LHC to account for both ATLAS and CMS accumulated
data\footnote{ It shall be mentioned that other analyses give
  more conservative prospects for the FCC-hh as well as for the
  HL-LHC, see for example ref.~\cite{Azatov:2015oxa}. However, new
  techniques in the meantime can be developed to help increasing the
  sensitivity, as well as a better analysis of possible search
  channels, see for example the case of the $4b$ final
  state~\cite{Behr:2015oqq}.}, as well as for the fact that the analysis in
ref.~\cite{He:2015spf} is only done for one search channel; we expect
the sensitivity to improve when more search channels are taken into
account. We display in Fig.~\ref{fig:sensitivity} the maximally
allowed deviation $\Delta^{\rm BSM}_{}$ (in percent), in black solid
line, as a function of the heavy neutrino rescaling factor $M_R^{}$
(in TeV). This is compared to the sensitivities to the SM prediction
for $\lambda_{HHH}^{}$ in the case of the HL-LHC with an integrated
luminosity of 3 ab$^{-1}_{}$ (dashed black line); the ILC with
different center-of-mass energies $\sqrt{s}$ and integrated
luminosities $\mathcal{L}$, $\sqrt{s}=500$~GeV and $\mathcal{L} = 4$
ab$^{-1}_{}$ (double dotted blue line), $\sqrt{s}=1$~TeV and
$\mathcal{L} =2$ ab$^{-1}_{}$ (dotted purple line), $\sqrt{s}=1$~TeV
and $\mathcal{L} = 5$ ab$^{-1}_{}$ (long dash-dotted green line); and
the case of the FCC-hh at 100 TeV and with $\mathcal{L} = 3$
ab$^{-1}_{}$ (dash-dotted red line). While the currently foreseen
sensitivity of the HL-LHC would not allow to resolve the effect of the
heavy neutrinos, new analysis techniques or the other future colliders
would clearly allow to test these heavy neutrino corrections.

More specifically and using current experimental constraints, the ILC
at a center-of-mass energy $\sqrt{s}=500$~GeV could probe heavy
neutrino masses in the range $8.5 < M_R^{} < 10.5$ TeV, at 1 TeV with
5 ab$^{-1}_{}$ of data this extends to the range $5 < M_R^{} < 17.5$
TeV. The FCC-hh collider could extend the analysis to a bigger range
$3.3 < M_R^{} < 20$~TeV. Indirect searches and in particular EWPO
could probe heavy neutrinos with masses in the multi-TeV range and
future improvements are expected, especially at future $e^+e^-$
colliders~\cite{Antusch:2015mia}. Improved constraints on the EWPO
would tend to shift the left-hand part of the black curve in
Fig.~\ref{fig:sensitivity} towards the right. This makes the triple
Higgs coupling a new, viable and attractive observable to test
low-scale seesaw mechanisms that will be complementary to improved
EWPO measurements.\bigskip

%
%
\section{Conclusions}
\label{sec:conc}

We have investigated in this article the one-loop effects of heavy
neutrinos on the triple Higgs coupling in the framework of an inverse
seesaw model, that is a realistic, renormalizable model accounting for
the masses and mixings of the light neutrinos.
After having presented the model and its constraints, both theoretical
and experimental, in Section~\ref{sec:model}, we have given the
technical details of the one-loop calculation in
Section~\ref{sec:calc}. We have presented in Section~\ref{sec:pheno}
our numerical investigation of the model. After having performed a
scan in a Casas-Ibarra parameterization of the neutrino input
parameters we have found that a $\mu_X^{}$-parameterization is more
suitable to get the maximal effects on the triple Higgs coupling and
we have obtained a deviation as high as $\sim +30\%$ for the class of
parameters in which the $3\times 3$ heavy neutrino mass matrix
$M_R^{}$ is diagonal and hierarchical while the $3\times 3$ neutrino
Yukawa texture is proportional to the identity matrix. This confirms
our expectations coming from the simplified model analysis, and
establishes the triple Higgs coupling as a viable, new observable to
probe heavy neutrino mass regimes that are hard to probe otherwise, as
this deviation is at the current limit for the expected sensitivity at
the HL-LHC but clearly visible at the ILC and at the FCC-hh. Heavy
neutrinos can also give rise to new diagrams that contribute to the
complete $HH$ production cross section and need to be evaluated. We
leave this for future projects.

%
%
\acknowledgments
We warmly thank Nad\`ege Bernard for her logistic support during the
last stage of the project. C.W. heartfully thanks the University of
T\"ubingen for its hospitality during the final stages of this
project. We also acknowledge the discussions with Juraj Streicher and
Bhupal Dev as well as with the participants of the \textit{Focus
  Meeting on Collider Phenomenology}, organized at the IBS CTPU,
Daejeon, South Korea. J.B. acknowledges the support from the
Institutional Strategy of the University of T\"ubingen ( DFG, ZUK 63)
and from the DFG Grant JA 1954/1. C.W. receives financial support from
the European Research Council under the European Union's Seventh
Framework Programme (FP/2007-2013)/ERC Grant NuMass Agreement
No.~617143 and partial support from the European Union's Horizon 2020
research and innovation programme under the Marie Sk\l{}odowska-Curie
grant agreements No.~690575 and No.~674896. The Feynman diagrams of
this article have been drawn with the program {\tt  JaxoDraw
  2.0}~\cite{Binosi:2003yf,Binosi:2008ig}.

%

\appendix

\section{Next order corrections in the seesaw expansion parameter to
  the \boldmath $\mu_X$-parameterization}
\label{app:NLO}

Following the method of ref.~\cite{Grimus:2000vj}, we can diagonalize
$M_{\mathrm{ISS}}$ by block to an arbitrary order in the seesaw
expansion parameter $m_D M_R^{-1}$. This gives for the $3\times 3$ light
neutrino mass matrix 
\begin{align}
  \label{NLOterms}
  \begin{split}
    M_{\mathrm{light}} =\,
    & m_D M_R^{T-1} \mu_X M_R^{-1} m_D^T -
    \frac{1}{2} m_D M_R^{T-1} M_R^{*-1}
    m_D^\dagger m_D M_R^{T-1} \mu_X M_R^{-1} m_D^T\\
    & -\frac{1}{2} m_D M_R^{T-1} \mu_X M_R^{-1} m_D^T m_D^*
    M_R^{\dagger-1} M_R^{-1} m_D^T + \mathrm{o}\left( ||M_R^{-1}
      m_D||^4 \right)\times \mu_X \,,
  \end{split}
\end{align}
in agreement with previous results~\cite{Hettmansperger:2011bt}. This
can be written in a symmetric form
\begin{align}
    M_{\mathrm{light}} =\,
    & m_D M_R^{T-1} \left(\mathbf{1}-\frac{1}{2} M_R^{*-1} m_D^\dagger
      m_D M_R^{T-1} \right) \mu_X \left(\mathbf{1}-\frac{1}{2} M_R^{-1}
      m_D^T m_D^* M_R^{\dagger-1}\right) M_R^{-1} m_D^T \nonumber\\
    & + \mathrm{o}\left( ||M_R^{-1} m_D||^4 \right)\times \mu_X \,.
\end{align}
If $m_D$ is invertible, we can then express $\mu_X$ as a function of
$M_{\mathrm{light}}$ and the other blocks of $M_{\mathrm{ISS}}$,
\begin{equation}
\label{nearlythere}
 \mu_X^{} \simeq \left(\mathbf{1}-\frac{1}{2} M_R^{*-1} m_D^\dagger
   m_D^{} M_R^{T-1} \right)^{-1}_{}\!\!\! M_R^T m_D^{-1}
 M_{\mathrm{light}}^{} m_D^{T-1} M_R^{} \left(\mathbf{1}-\frac{1}{2}
   M_R^{-1} m_D^T m_D^* M_R^{\dagger-1}\right)^{-1}_{}\,.
\end{equation}
The light neutrino mass matrix is diagonalized by using the unitary
PMNS matrix. Using eq.(\ref{mnulight}) to rewrite
$M_{\mathrm{light}}$ in eq.(\ref{nearlythere}), we get a formula for
the $\mu_X$-parameterization that includes the effect of sub-leading
terms in the seesaw expansion,
\begin{align}
  \label{app:muXparam}
  \begin{split}
    \mu_X \simeq
    & \left(\mathbf{1}-\frac{1}{2} M_R^{*-1} m_D^\dagger m_D M_R^{T-1}
    \right)^{-1} M_R^T m_D^{-1} U_{\rm PMNS}^* m_\nu U_{\rm
      PMNS}^\dagger m_D^{T-1} M_R\, \times\\
    & \left(\mathbf{1}-\frac{1}{2} M_R^{-1} m_D^T m_D^*
      M_R^{\dagger-1}\right)^{-1}\,.
  \end{split}
\end{align}
It is easy to see that if we were to consider only the leading order
term in the seesaw expansion, we would recover eq.(45)
from ref.~\cite{Arganda:2014dta}.

Interestingly, our results would not be modified by the addition of an
extra mass term $\mu_R \overline{\nu_{R}^C} \nu_{R}$. The neutrino
mass matrix would then be
\begin{equation}
  \label{ESSmatrix}
  M=\left(
    \begin{array}{c c c}
      0 & m_D & 0\\
      m_D^T & \mu_R & M_R \\
      0 & M_R^T & \mu_X
    \end{array}\right)\,,
\end{equation}
where taking $||\mu_R||\ll ||m_D||,||M_R||$ corresponds to the inverse
seesaw limit while taking $||\mu_R||\geq ||M_R||$ leads to the
extended seesaw limit. In both cases, the next order corrections to
$M_{\mathrm{light}}$ are given by eq.(\ref{NLOterms}) in the limit
where $||\mu_X M_R^{-1} \mu_R||\ll ||M_R||$. Thus, eq.(\ref{muXparam})
would remain unchanged\footnote{This conclusion is limited to the
  next-order term in the seesaw expansion. In general, one-loop
  corrections proportional to $\mu_R$ should also be included unless
  $\mu_R\ll\mu_X$ (see ref.~\cite{Dev:2012sg} and references
  therein).}.

\section{Analytic expressions of the new ISS contributions}
\label{app:HHH}

We give in this appendix all the analytic formulae of the new ISS
contributions involved in the calculation of the renormalized one-loop
triple Higgs coupling presented in Section~\ref{sec:calc}. The SM
contributions, denoted with a SM, can be found in
ref.~\cite{Denner:1991kt} and will not be reproduced in this appendix. 

\subsection{Counter-terms}

By convention all loop integrals in this sub-section are to be
understood as their real part only. We use the conventions of {\tt
  LoopTools 2.13}~\cite{vanOldenborgh:1990yc,Hahn:1998yk,Hahn:2006qw}
for the scalar integrals and the tensor coefficients.

\begin{align}
  \delta M_W^2 = \, & \delta M_W^2\big|_{\rm SM} -  \frac{\alpha}{4
                      \pi  s_W^2} 
                       \sum_{i=1}^3 \sum_{j=1}^9 \left| B_{i
                      j}\right|^2\bigg(A_0(m_{n_j}^2)+m_{\ell_i}^2
                      B_{0}(M_W^2,m_{\ell_i}^2,m_{n_j}^2)\nonumber\\
                    & -2 B_{00}(M_W^2,m_{\ell_i}^2,m_{n_j}^2)+M_W^2
                      B_{1}(M_W^2,m_{\ell_i}^2,m_{n_j}^2)\bigg)
\end{align}

\begin{align}
  \delta M_Z^2 = \, & \delta M_Z^2\big|_{\rm SM}- \frac{3
                      \alpha}{48\pi c_W^2 s_W^2} \sum_{j=1}^9
                      \sum_{k=1}^9 \bigg( \Big(C_{j k} C_{k j}^*+C_{j
                      k}^* C_{k j}\Big) m_{n_j} m_{n_k}
                      B_{0}(M_Z^2,m_{n_j}^2,m_{n_k}^2)\nonumber\\
                    & +\Big(C_{j k} C_{k j}+C_{j k}^* C_{k j}^*\Big)
                      \Big(A_{0}(m_{n_k}^2)+m_{n_j}^2
                      B_{0}(M_Z^2,m_{n_j}^2,m_{n_k}^2)-2
                      B_{00}(M_Z^2,m_{n_j}^2,m_{n_k}^2) \nonumber\\
                    & +M_Z^2
                      B_{1}(M_Z^2,m_{n_j}^2,m_{n_k}^2)\Big)\bigg)
\end{align}

\begin{align}
  \delta t_H =\, & \delta t_H\Big|_{\rm SM}
                   -\frac{\sqrt{2\pi\alpha}}{8\pi^2 M_W s_W}
                   \sum_{j=1}^9 m_{n_j}^2 \re(C_{ j j})
                   A_{0}(m_{n_j}^2)
\end{align}

\subsection{One-loop un-renormalized self energy $\Sigma_{HH}^{}$ and
  vertex $\lambda_{HHH}^{(1)}$}

The self-energy enters in the calculation of the field renormalization
as well as the Higgs mass $M_H^{}$ counter-term. In the one-loop un-renormalized triple
Higgs coupling, $q$ is the momentum of the off-shell Higgs boson splitting into
two Higgs bosons, $H^*_{}(q)\to H H$.

\begin{align}
  \Sigma_{HH}(p^2) = \, & \Sigma_{HH}^{\rm SM}(p^2) - \frac{\alpha
                          }{16\pi M_W^2  s_W^2} \sum_{j=1}^9
                          \sum_{k=1}^9 \bigg( \Big(C_{j k}
                          C_{k j}+C_{j k}^* C_{k j}^*\Big)
                          m_{n_j}^2 \Big(A_{0}(m_{n_k}^2)\nonumber\\
                        & +p^2
                          B_{1}(p^2,m_{n_j}^2,m_{n_k}^2)+m_{n_j}^2
                          B_{0}(p^2,m_{n_j}^2,m_{n_k}^2) \Big) +\Big(C_{j k}
                          C_{k j}+C_{j k}^* C_{k j}^*\Big) m_{n_k}^2
                          \Big(A_{0}(m_{n_k}^2)\nonumber\\
                        & +p^2 B_{1}(p^2,m_{n_j}^2,m_{n_k}^2) +3
                          m_{n_j}^2 B_{0}(p^2,m_{n_j}^2,m_{n_k}^2)
                          \Big) +
                          \Big(C_{k j} C_{j k}^*+C_{j k} C_{k j}^*\Big)
                          m_{n_j} m_{n_k}\nonumber\\
                        & \Big(2 A_{0}(m_{n_k}^2) +2 p^2
                          B_{1}(p^2,m_{n_j}^2,m_{n_k}^2)+3 m_{n_j}^2
                          B_{0}(p^2,m_{n_j}^2,m_{n_k}^2)
                          \Big) \nonumber\\
                        & +\Big(C_{k j} C_{j k}^*+C_{j k}
                          C_{k j}^*\Big) m_{n_j}
                          m_{n_k}^3
                          B_{0}(p^2,m_{n_j}^2,m_{n_k}^2)\bigg)
\end{align}

\begin{align}
  \allowdisplaybreaks
  \lambda^{(1)}_{HHH}(q) =\, & \lambda^{(1), {\rm SM}}_{HHH}(q) 
                      -\frac{\alpha \sqrt{4\pi\alpha}}{32 \pi M_W^3
                      s_W^3} \sum _{j=1}^9 \sum _{k=1}^9 \sum _{l=1}^9
                      \bigg[
                      \Big(C_{j k}^{} C_{k l}^{} C_{l j}^{}+C_{j k}^* C_{k l}^* C_{l j}^*\Big)\nonumber\\
                    &  \bigg(m_{n_j}^2 m_{n_k}^2 \left(4 B_0^{} + 4 M_H^2
                      C_2^{}+ q^2_{} (C_0^{} + 4 C_1^{} + C_2^{}) + 4
                      m_{n_j}^2 C_0^{}\right) \nonumber\\
                    & + m_{n_l}^2 m_{n_k}^2 \Big(4
                      B_0^{} + 2 M_H^2 C_2^{} + q^2_{} (3 C_1^{} +
                      C_2^{}) \Big) + m_{n_l}^2 m_{n_j}^2 \Big( 4 B_0^{} + \nonumber\\ 
                    &  
                      4 (m_{n_j}^2 + 2 m_{n_k}^2) C_0^{} + 2 M_H^2
                      C_2^{} + q^2_{} \left(C_0^{} + 5 C_1^{} + 2
                      C_2^{}\right) \Big) \bigg)+\nonumber\\
                    & 
                      m_{n_j} m_{n_l} \Big(C_{j k}^{} C_{k l}^{} C_{j
                      l}^{}+C_{j k}^* C_{k l}^* C_{j l}^*\Big) \bigg(
                      m_{n_l}^2 \Big(2 B_0^{} + q^2_{} \left(2 C_1^{} +
                      C_2^{} \right)\Big) +\nonumber\\
                    & m_{n_k}^2 \Big(8 B_0^{} + 6 M_H^2 C_2^{} +
                      q^2_{} (C_0^{} + 7 C_1^{} + 2 C_2^{}) + 2
                      m_{n_l}^2 C_0^{} \Big) + \nonumber\\ 
                    & m_{n_j}^2 \Big(2 B_0^{} + 2 M_H^2 C_2^{}
                      + q^2_{} (C_0^{} + 3 C_1^{} + C_2^{}) + 2 (5
                      m_{n_k}^2 + m_{n_j}^2 + m_{n_l}^2) C_0^{}
                      \Big)\bigg)+\nonumber\\
                    &
                      m_{n_k} m_{n_l}\Big(C_{j k}^{} C_{l k}^{} C_{l
                      j}^{}+C_{j k}^* C_{l k}^* C_{l j}^*\Big) \bigg(
                      m_{n_k}^2 \Big(2 B_0^{} + 2 M_H^2 C_2^{} +
                      q^2_{} C_1^{} \Big) + 
                      \nonumber\\ 
                    & m_{n_l}^2 \Big(2 B_0^{} + q^2_{} (2 C_1^{} +
                      C_2^{} ) \Big)
                      + m_{n_j}^2 \Big(8 B_0^{} + 6 M_H^2 C_2^{} 
                      + q^2_{} (2 C_0^{} + 9 C_1^{} + 3 C_2^{})
                      +\nonumber\\
                     & 4 \big(2 m_{n_j}^2 + m_{n_k}^2 +
                       m_{n_l}^2\big) C_0^{} \Big)\bigg)+\nonumber\\
                     & m_{n_j} m_{n_k}\Big(C_{j k}^{} C_{l k}^{} C_{j
                       l}^{}+C_{j k}^* C_{lk}^* C_{j l}^*\Big)
                     \bigg(m_{n_l}^2 \Big(8 B_0^{} + 4 M_H^2 C_2^{} +
                      q^2_{} (C_0^{} + 8 C_1^{} + 3 C_2^{}) \Big) +
                      \nonumber\\
                    & m_{n_k}^2 \Big(2 B_0^{} + 2 M_H^2 C_2^{} +
                      q^2_{} C_1^{} + 2 m_{n_l}^2 C_0^{}\Big) +
                      m_{n_j}^2 \Big(2 B_0^{} + 2 M_H^2 C_2^{} + \nonumber\\
                    & q^2_{} (C_0^{} + 3 C_1^{} + C_2^{}) + 2
                      C_0^{} (m_{n_j}^2 + m_{n_k}^2 + 5 m_{n_l}^2)\Big)\bigg)\bigg]
\end{align}

\noindent In the expression of the un-renormalized vertex
$\lambda_{HHH}^{(1)}$ we have used the following abbreviations, 
\begin{align}
B_0^{} & \equiv  B_0^{}\left(M_H^2,m_{n_k}^2,m_{n_l}^2\right)\,,\nonumber\\
C_0^{} & \equiv  C_0^{}\left(q_{}^2,M_H^2,M_H^2,
 m_{n_j}^2,m_{n_k}^2,m_{n_l}^2\right)\,,\nonumber\\
C_{1/2}^{} &  \equiv  C_{1/2}^{}\left(q_{}^2,M_H^2,M_H^2,
m_{n_j}^2,m_{n_k}^2,m_{n_l}^2\right)\, .
\end{align}

\bibliography{hhh_inverse-seesaw_jhep-v2}{}
\bibliographystyle{JHEP}

\end{document}